\documentstyle[11pt,epsfig]{article}
\begin{document}

\title{Dynamics of entanglement between quantum dot spin-qubits
\footnote{To appear in the Proceedings of the E. Fermi School,
"Quantum Phenomena of Mesoscopic Systems", 9 - 19 July 2002, Varenna, Italy.}}
\author{John Schliemann and Daniel Loss\\\\
Department of Physics and Astronomy, University of Basel,\\
CH-4056 Basel, Switzerland}

\date{\today}

\maketitle\abstract{We briefly review the physics of gate operations
between quantum dot spin-qubits and analyze the dynamics of quantum
entanglement in such processes. The indistinguishable
character of the electrons whose spins
realize the qubits gives rise to further entanglement-like quantum
correlations that go beyond simple antisymmetrization effects.
We also summarize further recent results concerning
this type of quantum correlations of indistinguishable particles.
Finally we discuss decoherence properties of spin-qubits when coupled to
surrounding nuclear spins in a semiconductor nanostructure}

\section{Introduction}

Quantum entanglement is one of the most intriguing features of quantum
mechanics. It is one of the decisive properties of the quantum world which
distinguishes it from classical physics. Moreover, in the last decade,
entanglement has proved to be the key ingredient of the emerging field
of quantum information theory\footnote{For general
introductions to the field of quantum information theory
see, e.g., \cite{Peres95,Steane98,Ekert00,Nielsen00,Werner01,Keyl02,Knill02},
for recent reviews on the issue of quantum entanglement see
\cite{Lewenstein00,Terhal02,Bruss02}.}.

In the beginning of modern quantum
theory, the notion of
entanglement was first noted by Einstein, Podolsky, and Rosen
\cite{Einstein35},
and by Schr\"odinger \cite{Schrodinger35}. While in those days quantum
entanglement and its predicted physical consequences were (at least partially)
considered as an unphysical property of the formalism (a ``paradox''),
the modern perspective on this issue is very different. Nowadays quantum
entanglement is to be seen as an experimentally verified property of nature
providing a resource for a vast variety of novel phenomena and concepts
such as quantum computation, quantum cryptography, or quantum teleportation.

While the basic notion of entanglement in pure quantum states of bipartite
systems (Alice and Bob) is theoretically well understood, fundamental
questions are open concerning entanglement in mixed states (described by
a proper density matrix),
or entanglement of more than two parties \cite{Lewenstein00,Terhal02,Bruss02}.
The most elementary example for entanglement in a pure quantum state is given
by a spin singlet composed from two spin-$\frac{1}{2}$-objects (qubits) owned
by A(lice) and B(ob), respectively,
\begin{equation}
\frac{1}{\sqrt{2}}\left(|\uparrow\rangle_{A}\otimes|\downarrow\rangle_{B}
-|\downarrow\rangle_{A}\otimes|\uparrow\rangle_{B}\right)\,.
\end{equation}
For such a state, the state of the combined system cannot be described by
specifying the state of Alice's and Bob's qubit separately. It is a standard
result of quantum information theory \cite{Peres95,Nielsen00} that this
property does not
depend on the basis chosen in Alice's or Bob's Hilbert space.
As we shall see below, the entanglement
of such a quantum state (quantified by an appropriate measure)
is invariant under (independent) changes of basis in both spaces.

Physically measurable consequences of quantum entanglement of the
above kind arise typically (but not exclusively) in terms of two-body
correlations between the subsystems. In this case the effects of
entanglement can typically be cast in terms of so-called Bell
inequalities \cite{Bell64} whose violation manifests the presence of
entanglement in a given quantum state. Using this formal approach
the physical existence of quantum entanglement (as opposed to classical
correlations) has unambiguously been verified by Aspect and coworkers
for the polarization state of photons\cite{Aspect82}.
Moreover, quantum entanglement is an essential ingredient of
algorithms for quantum computation \cite{Steane98,Ekert00,Nielsen00},
in particular for Shor's algorithm for decomposing large numbers into their
prime factors \cite{Shor94}. This problem is intimately related to public key
cryptography systems such as RSA encoding which is widely used in today's
electronic communication.

Among the many proposals for experimental realizations of quantum information
processing solid state systems have the advantage of offering the
perspective of scalability, i.e. the integration of a large number of quantum gates
into a quantum
computer once the single gates and qubits are established.
In the recent years, several proposals for using spins of electrons and/or
nuclei in solid state systems have been put forward, starting with the work
of Ref.~\cite{Loss98}, see Refs.~\cite{Privman98,Kane98,Barnes00,Levy01,Ladd01}.
Specifically, in  Ref.~\cite{Loss98} it was proposed to use the spin of
electrons residing in semiconductor quantum dots as qubits
\cite{Burkard99,Hu00,Schliemann01a,Friesen02,Vandersypen02}. The central idea is
to store the information in the spin-degree-of freedom of the electron, while the manipulation
of the spin, like single spin rotations and spin-spin interactions, are
achieved via electrical gates acting on the charge degree-of-freedom
of the electron.
In this contribution we briefly review essential aspects of this proposal
and analyze the dynamics of quantum entanglement in gate operations
between qubits of this type.
During such processes, the indistinguishable character of the electrons
leads to entanglement-like quantum correlations which require a
description different from the usual entanglement between distinguishable
parties (Alice, Bob, ...) in bipartite (or multipartite) systems.
In such a case the proper statistics of the indistinguishable particles has
to be taken into account \cite{Schliemann01a}.

This article is organized as follows. In section \ref{disting} we give a
brief overview on elementary results on entanglement
between {\em distinguishable} parties. In section \ref{dots} we briefly
review essential aspects of the quantum dot spin qubit proposal of
Ref.~\cite{Loss98}. After some general remarks we summarize in subsection
\ref{decoherence} recent results on decoherence properties of electron spins
interacting with surrounding nuclear spins in semiconductor nanostructures
\cite{Khaetskii02,Schliemann02}.
In subsection \ref{model} we give a brief description of the
low-energy physics of lateral quantum dot molecules consisting of two
spin-qubits.
This discussion will lead us to the question of
``entanglement'' (or entanglement-analogous quantum correlations)
between {\em indistinguishable} particles, which are in our case here
electrons in the joint system of two (or more) quantum dots
\cite{Schliemann01a}. The research and literature
on quantum entanglement has so far fairly concentrated on distinguishable
parties; the issue of indistinguishable (or, in other words, identical)
objects has been addressed only very recently
\cite{Schliemann01a,Schliemann01b,Eckert02,Paskauskas01,Li01,Zanardi01,Zanardi02,Omar02,Gittings02,Ghirardi02,Zeng02,Wiseman02,Martin-Delgado02}.
We therefore give also a
short overview on this new direction of work in quantum information theory.
In section \ref{dynamics}
we discuss results on gate operations between quantum dot
spin qubits with the emphasis on the {\em dynamics of quantum entanglement}.
Specifically we present results on swap processes (interchanging
the state of two qubits) and the ``square root of a swap'' generating
a maximally entangled state from a tensor product.
We close with a summary in section \ref{summary}.


\section{Quantum Entanglement between distinguishable parties}
\label{disting}

We now give an introduction to basic concepts of characterizing and
quantifying entanglement between distinguishable parties. We
concentrate on pure states (i.e. elements of the joint Hilbert space)
of bipartite systems. We then comment only briefly on the case of mixed states
(described by a proper density operator), and entanglement in multipartite
systems.

One of the most prominent examples of an {\em entangled state} was
already given in the previous section, namely a spin singlet built up
from two qubits. More generally, if Alice and Bob own Hilbert spaces
${\cal H}_{A}$ and ${\cal H}_{B}$ with dimensions $m$ and $n$, respectively,
a state $|\psi\rangle$ is called {\em nonentangled} if it can be
written as a product state,
\begin{equation}
|\psi\rangle=|\alpha\rangle_{A}\otimes|\beta\rangle_{B}
\label{prodstate}
\end{equation}
with $|\alpha\rangle_{A}\in{\cal H}_{A}$, $|\beta\rangle_{B}\in{\cal H}_{B}$.
Otherwise $|\psi\rangle$ is entangled. The question arises whether a
given state $|\psi\rangle$, expressed in some arbitrary basis of
the joint Hilbert space ${\cal H}={\cal H}_{A}\otimes{\cal H}_{B}$,
is entangled or not, i.e. whether there are states
$|\alpha\rangle_{A}$ and $|\beta\rangle_{B}$
fulfilling (\ref{prodstate}). Moreover, one would like to quantify the
entanglement contained in a state vector.

An important tool to investigate such questions for bipartite systems is the
biorthogonal Schmidt decomposition \cite{Peres95}.
It states that for any state vector $|\psi\rangle\in{\cal H}$
there exist bases of ${\cal H}_{A}$ and ${\cal H}_{B}$ such that
\begin{equation}
|\psi\rangle=\sum_{i=1}^{r}z_{i}
\left(|a_{i}\rangle\otimes|b_{i}\rangle\right)
\label{Schmidt}
\end{equation}
with coefficients $z_{i}\neq 0$ and the basis states fulfilling
$\langle a_{i}|a_{j}\rangle=\langle b_{i}|b_{j}\rangle=\delta_{ij}$.
Thus, each vector in both bases for ${\cal H}_{A}$ and ${\cal H}_{B}$
enters at most only one product vector in the above expansion.
As a usual convention, the phases of the basis vectors involved in
(\ref{Schmidt}) can be chosen such that all $z_{i}$ are positive.
The expression (\ref{Schmidt}) is an expansion of the state $|\psi\rangle$
into a basis of product vectors $|a\rangle\otimes|b\rangle$ with a minimum
number $r$ of nonzero terms. This number ranges from $1$ to
$\min\{m,n\}$ and is called the {\em Schmidt rank} of $|\psi\rangle$.

With respect to arbitrary bases in  ${\cal H}_{A}$ and ${\cal H}_{B}$ a
given state vector reads
\begin{equation}
|\psi\rangle=\sum_{a,b}M_{ab}|a\rangle\otimes|b\rangle
\end{equation}
with an $m\times n$ coefficient matrix $M$. Under unitary transformations
$U_{A}$ and $U_{B}$ in ${\cal H}_{A}$ and ${\cal H}_{B}$, respectively, $M$
transforms as
\begin{equation}
M\mapsto M'=U_{A}MU_{B}^{T}
\end{equation}
with $U_{B}^{T}$ being the transpose of $U_{B}$. The fact that there are always
bases in ${\cal H}_{A}$ and ${\cal H}_{B}$ providing a biorthogonal
Schmidt decomposition of $|\psi\rangle$ is equivalent to stating that
there are matrices $U_{A}$ and $U_{B}$ such that the resulting matrix
$M'$ consists of a diagonal block with only nonnegative entries while the
rest of the matrix contains only zeros. For the case of equal dimensions
of Alice's and Bob's space, $m=n$, this is also a well-known theorem of
matrix algebra \cite{Mehta77}.

Obviously, $|\psi\rangle$ is nonentangled, i.e. a simple product state,
if and only if its Schmidt rank is one. More generally, the Schmidt rank of
a pure state can be viewed as a rough characterization for its entanglement.
However, since the Schmidt rank is by construction a discrete quantity
it does not provide a proper quantification of entanglement. Therefore
finer entanglement measures are desirable.
For the case of two distinguishable parties, a useful measure of
entanglement is the von Neumann-entropy of partial density matrices
constructed from the pure-state density matrix $\rho=|\psi\rangle\langle\psi|$
\cite{Bennett96}:
\begin{equation}
E(|\psi\rangle)=-{\rm tr}_{A}\left(\rho_{A}\log_{2}\rho_{A}\right)
=-{\rm tr}_{B}\left(\rho_{B}\log_{2}\rho_{B}\right)\,,
\label{vNeuEnt}
\end{equation}
where the partial density matrices are obtained by tracing out one of the
subsystems, $\rho_{A/B}={\rm tr}_{B/A}\rho$. With the help of the
biorthogonal Schmidt decomposition of $|\psi\rangle$ one shows that
both partial density matrices have the same spectrum and therefore the
same entropy, as stated in Eq.~(\ref{vNeuEnt}). In particular, the
Schmidt rank of $\psi\rangle$ equals the algebraic rank of the partial
density matrices. $|\psi\rangle$ is nonentangled if and only if the
partial density matrices of the pure state $\rho=|\psi\rangle\langle\psi|$
are also pure states, and $|\psi\rangle$ is maximally entangled if its
partial density matrices are ``maximally mixed'', i.e. if they have only
one non-zero eigenvalue with a multiplicity of $\min\{m,n\}$.

It is important to observe that the entanglement
measure (\ref{vNeuEnt}) of a given state $|\psi\rangle$
does not depend on the bases used in Alice's and Bob's Hilbert space
to express this state.
This is because the trace operations in the definition of
$E(|\psi\rangle)$ are invariant under a change
of bases (performed, in general, independently in both spaces).
Therefore, entanglement in bipartite systems is a basis-independent
quantity.

Thus, the problem of characterizing and quantifying
quantum entanglement for pure states in bipartite systems can been
seen as completely solved. Unfortunately, the situation is much less clear
for mixed states \cite{Lewenstein00,Terhal02,Bruss02},
and for multipartite entanglement. The main obstacle in the latter issue
is the fact that the biorthogonal Schmidt
decomposition in bipartite systems does not have a true analogue
in the multipartite case.


\section{Quantum entanglement with electron spins in quantum dots}
\label{dots}

We will now illustrate the phenomenon of quantum entanglement on the
example of a specific (possible) realization of a quantum information
processing system \cite{Loss98}. The proposal to be discussed below
deals with qubits realized by the spins of electrons residing on
semiconductor quantum dots. As we shall see in this and the following
section, the indistinguishable character of the electrons gives rise
to quantum correlations which are beyond entanglement between distinguishable
parties.

An study analogous to the one to be described below \cite{Schliemann01a}
has been carried out recently for the case of indistinguishable bosons
(instead of electrons, i.e.  indistinguishable fermions)
in Refs.~\cite{Mompart02}. There the quantum dynamics (and generation of
entanglement) between bosonic atoms trapped in neighboring potential
minima of an optical lattice was studied and numerically simulated.

\subsection{General remarks}

An array of coupled quantum dots, see figure \ref{figArray},
each dot containing
a top most spin 1/2, was found to be a promising candidate for a scalable
quantum computer~\cite{Loss98} where the qubit is
defined by the spin $1/2$  on the dot. Quantum algorithms can then be
implemented using
local single-spin rotations and the exchange coupling between nearby spins,
see figure \ref{figArray}. This proposal is supported by experiments where,
e.g., Coulomb blockade effects,~\cite{Waugh96}
tunneling between neighboring dots,~\cite{Kouwenhoven97,Waugh96}
and magnetization~\cite{Oosterkamp98} have been observed as well as the
formation of a delocalized single-particle state in coupled dots
\cite{Blick98}.

The basic mechanism
for two-qubit gate operations within quantum-dot spin-qubit proposal
are gated (i.e. time-dependent) tunneling amplitudes between the dots.
Then the two-qubit dynamics is driven by the effective exchange coupling
between the electron spins on different dots arising from a finite tunneling
amplitude.

In the following subsection
we shall briefly address the issue of decoherence of spin-qubits.
In subsection~\ref{model} we will summarize properties of a specific physical
description of quantum dot spin-qubits which are necessary for our
discussion here. For more comprehensive reviews of quantum computing
with electron spins in quantum dots see
Refs.~\cite{Golovach02,Gywat02,Burkard02}.

\subsection{Decoherence properties}
\label{decoherence}

The electron spin is a natural candidate for a qubit since its spin state in
a given direction, $|\uparrow\rangle$ or $|\downarrow\rangle$,
can be identified with the classical
bits $|0\rangle$ and $|1\rangle$, while an arbitrary superposition
$\alpha|\uparrow\rangle+\beta|\downarrow\rangle$ defines a qubit.
In principle, any quantum two-level system can be used to define a qubit.
However, one must be able to control coherent superpositions of
the basis states of the quantum computer,
i.e. no transition from quantum to classical behavior should occur.
Thus, the coupling of the environment to the qubit should be small,
resulting in a sufficiently large decoherence time $T_2$
(the time over which the phase of a superposition of $|0\rangle$
and  $|1\rangle$ is well-defined).
Assuming weak spin-orbit effects,
the spin decoherence time $T_{2}$ can be completely different
from the charge decoherence time (a few nanoseconds),
and in fact it is known~\cite{Kikkawa97} that $T_{2}$ can be orders of
magnitude longer than nanoseconds.
Time-resolved optical measurements were used to determine $T_2^{*}$,
the decoherence time of an ensemble of spins, with $T_2^{*}$
exceeding 100 ns in bulk GaAs \cite{Kikkawa97}.
More recently, the single spin relaxation time $T_1$
(usually $T_1 \geq T_2$) of a single quantum dot attached to leads
was measured via transport to be longer than a few $\mu$s \cite{Fujisawa01},
consistent with calculations \cite{Khaetskii00}.

A particularly relevant and,
from a fundamental point of view, also interesting mechanism for electron spin
decoherence in quantum dots (or other semiconductor nanostrcutures)
is the hyperfine interaction with surrounding nuclear spins, which are
ubiquitous in many materials such as GaAs. Thus, the Hamiltonian reads
\begin{equation}
{\cal H}=\vec S\cdot\sum_{i}A_{i}\vec I_{i}
\label{defham}
\end{equation}
where $\vec S$ is the electron spin and $\vec I_{i}$ are the surrounding
nuclear spins (both taken here to be dimensionless)
The coupling constants $A_{i}$ have dimension of energy and are proportional
to the square modulus of the electron wave function at the location of the
$i$-th nucleus. Since the electron wave function is spatially varying
across the dot (a typical typical form for such envelope wave functions is
given by a Gaussian), the coupling is {\em inhomogeneous}, i.e. different
nuclei couple with different strength to the electron spin.
This system was investigated in detail very recently in
Refs.~\cite{Khaetskii02,Schliemann02}. Generally, the quantum dynamics
according to (\ref{defham}) lead to a {\em decay} of the electron spin
as measured in terms of its expectation values $\langle\vec S\rangle$
(assuming an initially fully polarized spin, $|\langle\vec S\rangle|=1/2$).
This spin decay is generally a source of decoherence, i.e. loss
of quantum information encoded in the qubit.
However, a main finding in
Ref.~\cite{Khaetskii02} is an exact analytical solution for the
quantum dynamics for the situation of an initially fully polarized nuclear
spin system with the electron spin pointing opposite to it. This exact
solution can be evaluated explicitly in the  limit of a large number
$N$ of nuclear spins being involved. As a result, the decay of the
electron spin is in this case only of order $1/N$ and occurs {\em after}
a time interval of order $\hbar N/A$, where $A=\sum_{i}A_{i}\propto
10^{-5}{\rm eV}$ for GaAs. In a typical GaAs quantum dot $N$ is of order
$10^{5}$. Therefore, for a sufficiently large number of nuclear spins being
involved, the magnitude of the electron spin decay is small,
and large decoherence times can be obtained.

The findings of Ref.~\cite{Khaetskii02}
have been corroborated and extended by numerical work in
Ref.~\cite{Schliemann02}. These investigations revealed a striking dependence
of the electron spin dynamics on the character of the initial state
of the nuclear spin system (having a general degree of polarization):
If the initial state of the total system is a simple tensor product
of the states each spin (both electron and nuclei), the time evolution
depends strongly on the initial nuclear spin state.
However, this behavior is very different from the situation where
the electron is initially still uncorrelated with the nuclei,
but the nuclei are among themselves {\em randomly correlated}, i.e.
in a random superposition of all tensor product basis states of the
nuclear system. Then the time evolution is very
{\em reproducible}, i.e. it does
(almost) not depend on the specific randomly generated initial state of the
nuclear system; examples of such observations are shown
in figure \ref{nuclearfig1}.
Moreover, the time evolution of such a randomly correlated
initial state mimics closely the average over the dynamics of all
possible tensor product initial states. This observation can be understood as
a {\em self-averaging property} \cite{Schliemann02}.
The above results are illustrated in figures \ref{nuclearfig1} and
\ref{nuclearfig2}

Another observation from the numerical studies in Ref.~\cite{Schliemann02}
is the  that the spin decay occurs more slowly if the
nuclear spins are initially in a tensor product (i.e. non-entangled) state.
This result can be expected to be of quite general nature and independent of
the specific model (\ref{defham}): {\em Disentangling the environment
suppresses decoherence}.

The decay of the central electron spin $\vec S(t)$ in (\ref{defham})
is generally accompanied with the generation of {\em entanglement} between the
electron and the nuclei \cite{Schliemann02}.
This entanglement can be measured conveniently by
the von-Neumann entropy of the
partial density matrix where either the electron or the
environment has been traced out from the pure-state density matrix
of the total state \cite{Bennett96}, cf. Eq.~(\ref{vNeuEnt}). Tracing out the
nuclear system we have
\begin{equation}
\rho_{el}(t)=\left(
\begin{array}{cc}
\frac{1}{2}+\langle S^{z}(t)\rangle & \langle S^{+}(t)\rangle\\
\langle S^{-}(t)\rangle & \frac{1}{2}-\langle S^{z}(t)\rangle
\end{array}
\right) \, .
\end{equation}
This matrix has eigenvalues $\lambda_{\pm}=1/2\pm|\langle\vec
S(t)\rangle|$, and the measure of entanglement reads
$E=-\lambda_{+}\log\lambda_{+}-\lambda_{-}\log\lambda_{-}$. Thus, the
formation of expectation values $|\langle\vec S(t)\rangle|<1/2$
is a manifestation of the entanglement
between the electron spin and the nuclear spin system. The maximum
entanglement, $E=\log 2$, is achieved if the electron spin has
decayed completely as measured by the expectation values of its
components, $\langle\vec S(t)\rangle=0$. The generation of quantum
entanglement between the electron spin and the nuclear spin system
signaled by a reduced value of $\langle\vec S(t)\rangle$ is a
main and crucial difference between the quantum system studied
here and its classical 'counterpart' described by a system of
Landau-Lifshitz equations.

\subsection{Modeling of the double quantum dot system}
\label{model}

Let us now consider a system of two laterally tunnel-coupled dots
having one electron each. Using an appropriate model
\cite{Burkard99,Hu00,Schliemann01a}
theoretical calculations have demonstrated the possibility of performing
two-qubit quantum gate operations in such a system by varying the
tunnel barrier between the dots. An important point to observe here
is the fact that the electrons whose spins realize the qubits are
indistinguishable particles \cite{Schliemann01a}. Differently from
the usual scenario of distinguishable parties (Alice, Bob, ...)
the proper quantum statistics has to be taken into account when a finite
tunneling between the dots is present
\cite{Schliemann01a,Schliemann01b,Eckert02}.

The Hamiltonian of the double dot system
is given by ${\cal H}=T+C$, where $C$ denotes the Coulomb
repulsion between the electrons, and $T=\sum_{i=1,2}h_{i}$ is the
single-particle part with
\begin{equation}
h_{i}=\frac{1}{2m}\left(\vec p_{i}+\frac{e}{c}\vec A(\vec r_{i})\right)^{2}
+V(\vec r_{i})\,.
\label{opham}
\end{equation}
The single-particle Hamiltonian $h_{i}$ describes electron dynamics confined
to the $xy$-plane in a perpendicular magnetic field B. The effective mass
$m$ is a material-dependent parameter.
The coupling of the dots (which includes tunneling)
is modeled by a quartic potential
\begin{equation}
V(x,y)=\frac{m\omega^{2}_{0}}{2}
\left(\frac{1}{4a^2}\left(x^{2}-a^{2}\right)^{2} +y^{2}\right)\,,
\end{equation}
which separates into two harmonic wells of frequency $\omega_{0}$ (one for
each dot) in the limit $2a\gg 2a_{0}$, where $a$ is half the distance between
the dots and $a_{0}=\sqrt{\hbar/m\omega_{0}}$ is the effective Bohr radius
of a dot.

Following Burkard {\it et al.} \cite{Burkard99} we employ the Hund-Mulliken
method of molecular orbits to describe the low-lying spectrum of our system.
This approach concentrates on the lowest orbital states in each dot and is
an extension of the Heitler-London method also discussed in \cite{Burkard99}.
The Hund-Mulliken approach accounts for
the fact that both electrons can, in the presence of a finite tunneling
amplitude, explore the entire system of the two dots. Therefore this
approach is suited to investigate the issue of entanglement-analogous
quantum correlations between indistinguishable particles.
In particular, the Hund-Mulliken approach includes two-particle states
with both electrons being on the same dot. This issue of
{\em double occupancies} will be discussed in more detail below.

In the usual symmetric gauge $\vec A=B(-y,x,0)/2$ the Fock-Darwin ground state
of a single dot with harmonic confinement centered around $\vec r=(\pm a,0,0)$
reads
\begin{eqnarray}
\varphi_{\pm a}(x,y) & = & \sqrt{\frac{m\omega}{\pi\hbar}}
\exp\left(\frac{m\omega}{2\hbar}\left(\left(x\mp a\right)^{2}+y^{2}\right)
\right)\nonumber\\
 & & \cdot\exp\left(\pm\frac{i}{2}y\frac{a}{l_{B}^{2}}\right)\,,
\label{dotstates}
\end{eqnarray}
where $l_{B}=\sqrt{\hbar c/eB}$ is the magnetic length, and the frequency
$\omega$ is given by $\omega^{2}=\omega_{0}^{2}+\omega_{L}^{2}$ where
$\omega_{L}=eB/2mc$ is the usual Larmor frequency. From these non-orthogonal
single-particle states we construct the orthonormalized states $|A\rangle$ and
$|B\rangle$ with wave functions
\begin{eqnarray}
\langle\vec r|A\rangle & = &
\frac{1}{\sqrt{1-2Sg-g^{2}}}\left(\varphi_{+a}-g\varphi_{-a}\right)\,,\\
\langle\vec r|B\rangle & = &
\frac{1}{\sqrt{1-2Sg-g^{2}}}\left(\varphi_{-a}-g\varphi_{+a}\right)\,,
\end{eqnarray}
with $S$ being the overlap between the states (\ref{dotstates}) and
$g=(1-\sqrt{1-S^{2}})/S$.
For appropriate values of system parameters such as the interdot distance
and the external magnetic field, the overlap $S$ becomes exponentially small
\cite{Burkard99}. In this limit an electron in one of the states
$|A\rangle$, $|B\rangle$ is predominantly localized around
$\vec r=(\pm a,0,0)$.
In the following we consider this case and use these states as basis states
to define qubits, i.e.
qubits are realized by the spin state of an electron in either orbital
$|A\rangle$, or orbital $|B\rangle$.

An appropriate basis set for the six-dimensional two-particle Hilbert space is
given (using standard notation) by the three spin singlets
\begin{eqnarray}
|S_{1}\rangle & = & \frac{1}{\sqrt{2}}
\left(c^{+}_{A\uparrow}c^{+}_{B\downarrow}-
c^{+}_{A\downarrow}c^{+}_{B\uparrow}\right)|0\rangle\,,\\
|S_{2}\rangle & = & \frac{1}{\sqrt{2}}
\left(c^{+}_{A\uparrow}c^{+}_{A\downarrow}+
c^{+}_{B\uparrow}c^{+}_{B\downarrow}\right)|0\rangle\,,\\
|S_{3}\rangle & = & \frac{1}{\sqrt{2}}
\left(c^{+}_{A\uparrow}c^{+}_{A\downarrow}-
c^{+}_{B\uparrow}c^{+}_{B\downarrow}\right)|0\rangle\,,
\end{eqnarray}
and the triplet multiplet,
\begin{eqnarray}
|T^{-1}\rangle & = &
c^{+}_{A\downarrow}c^{+}_{B\downarrow}|0\rangle\,,\\
|T^{0}\rangle & = & \frac{1}{\sqrt{2}}
\left(c^{+}_{A\uparrow}c^{+}_{B\downarrow}+
c^{+}_{A\downarrow}c^{+}_{B\uparrow}\right)|0\rangle\,,\\
|T^{1}\rangle & = & c^{+}_{A\uparrow}c^{+}_{B\uparrow}|0\rangle\,.
\end{eqnarray}
The three triplet states are  degenerate (typically we can ignore possible
Zeeman splittings \cite{Burkard99}) and have the
common  eigenvalue,
\begin{equation}
\varepsilon_{T}=2\varepsilon+V_{-}\,,
\end{equation}
where we have defined
\begin{equation}
\varepsilon=\langle A|h|A\rangle=\langle B|h|B\rangle
\end{equation}
and
\begin{equation}
V_{-}=\langle T^{\alpha}|C|T^{\alpha}\rangle\quad,\quad
V_{+}=\langle S_{1}|C|S_{1}\rangle\,.
\end{equation}

An important further observation is that, as a consequence of inversion
symmetry along the axis connecting the dots,
the Hamiltonian does not
have any non-zero matrix elements between the singlet state $|S_{3}\rangle$
and other states. Hence, $|S_{3}\rangle$ is, independently of
the system parameters, an eigenstate. The eigenvalues of the triplet and
$|S_{3}\rangle$, however, do depend on system parameters. The
Hamiltonian acting on the remaining space spanned by $|S_{1}\rangle$ and
$|S_{2}\rangle$ can be written as
\begin{equation}
{\cal H}=2\varepsilon+\frac{1}{2}U_{H}+V_{+}-\left(
\begin{array}{cc}
 U_{H}/2 & 2t_{H} \\ 2t_{H} & -U_{H}/2
\end{array}\right)
\label{2ham}
\end{equation}
where
\begin{equation}
t_{H}=-\langle A|h|B\rangle-\frac{1}{2}\langle S_{2}|C|S_{1}\rangle
\end{equation}
and
\begin{equation}
U_{H}=\langle S_{2}|C|S_{2}\rangle-V_{+}\,.
\end{equation}
The nontrivial part of (\ref{2ham}) is a simple Hubbard Hamiltonian
and can be identified as the Hamiltonian of a
pseudospin-$\frac{1}{2}$-object in a
pseudomagnetic field having a component $U_{H}$ in the $\hat z$-direction and
$4t_{H}$ in
the $\hat x$-direction of pseudospin space.  [Note that this
pseudospin is not related to the spin degree of freedom which provides the
qubit!]
The space spanned by $|S_{1}\rangle$ and
$|S_{2}\rangle$ contains the ground state of the system.
The basis states
themselves are eigenstates only in the case of vanishing tunneling amplitude
$t_{H}$ where $|S_{1}\rangle$ is the ground state.
In all other cases, the ground state has an admixture of doubly
occupied states contained in $|S_{2}\rangle$.
The energy gap between the triplet and the singlet ground state is
\begin{equation}
\varepsilon_{T}-\varepsilon_{S0}=V_{+}-V_{-}-\frac{U_{H}}{2}+
\frac{1}{2}\sqrt{U^{2}_{H}+16t^{2}_{H}}\,.
\label{stsplitting}
\end{equation}

A key challenge is the construction of systems composed of
two coupled quantum dots which can be coupled to perform
swap operations ${\cal U}_{SW}$, i.e. unitary two-qubit operations
which interchange the spin states (qubits) of the
electrons on the two dots \
By combining the ``square root'' ${\cal U}_{SW}^{1/2}$
of such a swap with other isolated-qubit manipulations one can construct a quantum
XOR gate. A quantum XOR gate, along with isolated-qubit operations, has been
shown to be sufficient for the implementation of any quantum algorithm
\cite{DiVincenzo95}. Hence
a practical and reliable realization of a swap gate would be an important
step toward the fabrication of a solid state quantum computer.
A swap operation in the present system is a unitary transformation which turns
a state having the qubits in different states, say,
\begin{equation}
c^{+}_{A\uparrow}c^{+}_{B\downarrow}|0\rangle=
\frac{1}{\sqrt{2}}\left(|T^{0}\rangle+|S_{1}\rangle\right)\,,
\label{instate}
\end{equation}
into a state where the contents of the qubits is interchanged,
\begin{equation}
c^{+}_{A\downarrow}c^{+}_{B\uparrow}|0\rangle=
\frac{1}{\sqrt{2}}\left(|T^{0}\rangle-|S_{1}\rangle\right)\,.
\label{outstate}
\end{equation}
These two states are eigenstates in the case $V_{+}=V_{-}$ and $t_{H}=0$
for which the singlet-triplet splitting vanishes.

As discussed in references \cite{Loss98,Burkard99,Schliemann01a},
swapping may be achieved
by the action of a gate that lowers the potential barrier between the
quantum dots.  This
leads to exponentially larger values for both $V_{+}-V_{-}$ and
$t_H$.  It is adequate for our purposes to consider a model where
$V_{+}=V_{-}$ (consistent with the above limit of small overlap $S$), and
the singlet-triplet splitting results entirely from
$t_H$.  If the duration and amplitude of a tunneling pulse
is adjusted
appropriately, the relative phase between the singlet and the triplet state
involved picks up a shift of $\pi$, and a swapping operation is performed.

A finite tunneling
amplitude necessarily leads to a finite probability for double occupancies of
qubit states where both electrons are on the same dot.
If double occupancy errors occur to any sizable extent
{\em as a result} of the
swapping process, any quantum computation based on this hardware is likely to
fail.
However, if the double occupancies are sufficiently
rare {\em after} the swapping process, errors in the quantum computation can
likely
be corrected dynamically.
An important observation is that the double-occupancy probability
{\em after} the swap vanishes in the adiabatic limit, {i.e.} if the
ramp time $\tau$ of the quantum gate is such that
$\hbar /\tau $ is much larger than the pseudospin splitting
$\sqrt{U_H^{2}+16t_H^2}$.
This follows since the non-adiabatic effects can arise only from
the states $|S_{1}\rangle$ and $|S_{2}\rangle$, which have a non-trivial
time evolution when the tunneling amplitude $t_{H}$ is time-dependent.
Thus, the question
of whether double occupancies are problematic for swap operations in the
present system is reduced to the question of how close the motion of
a spin-$\frac{1}{2}$-object in a time-dependent magnetic field is to its
adiabatic limit. This will be investigated further in Sec. \ref{dynamics}.

The reduction of the dynamics to the time evolution of a two-level-system
relies on the fact that the system has inversion symmetry along the
$\hat x$-axis in real space connecting the dots. This symmetry can be
broken if odd powers of the particle coordinates $x_{i}$ are added to
the Hamiltonian (\ref{opham}) like for example the potential of a
homogeneous electric field. However, the only additional matrix element
due to such terms in the Hamiltonian occurs in the subspace of double
occupied states between the singlets $|S_{2}\rangle$ and
$|S_{3}\rangle$. Thus, in the presence of an electric field
${\cal E}=-eE\sum_{i}x_{i}$ the Hamitonian acting on the singlet
subspace spanned by $|S_{1}\rangle$, $|S_{2}\rangle$, $|S_{3}\rangle$ reads
\begin{eqnarray}
{\cal H} & = & 2\varepsilon+\frac{1}{2}U_{H}+V_{+}\nonumber\\
 & & -\left(
\begin{array}{ccc}
 U_{H}/2 & 2t_{H}  & 0 \\
 2t_{H} & -U_{H}/2 & F \\
 0 & F & -U_{H}/2+2X
\end{array}\right)
\label{3ham}
\end{eqnarray}
with the real matrix element $F=\langle S_{2}|{\cal E}|S_{3}\rangle$ and
\begin{equation}
2X=\langle S_{2}|C|S_{2}\rangle-\langle S_{3}|C|S_{3}\rangle
=2\langle A|\langle A|C|B\rangle|B\rangle\,.
\end{equation}
With a finite matrix element $F$ the dynamics of the system
is slightly more complicated, but also in this case the only coupling
of the two-qubit states (\ref{instate}) and (\ref{outstate}) to the
subspace of double occupied states is
provided by the tunneling amplitude $t_{H}$. Therefore, with respect
to the adiabaticity of the swapping process, the situation can be
expected to be not very different from the one with inversion symmetry
between the dots. This will be verified in Sec. {\ref{dynamics}.

So far we have not considered a possible Zeeman coupling to the electron spin.
This would not change the situation essentially since all states
involved in the
swapping  process ($|T^{0}\rangle$, $|S_{1}\rangle$, $|S_{2}\rangle$, and
possibly $|S_{3}\rangle$) have the
total spin quantum number $S^{z}=0$.

In the following section we give an elementary introduction to
the theory of ``entanglement-like'' quantum correlations in systems of
indistinguishable particles. We concentrate on the fermionic case and
illustrate our findings on the above example of coupled quantum dots.


\section{Quantum Correlations between indistinguishable particles}
\label{indisting}

For indistinguishable particles a pure
quantum state must be formulated in terms of Slater determinants or
Slater permanents for fermions and bosons, respectively.
Generically, a Slater determinant contains correlations due to the exchange
statistics of the indistinguishable fermions. As the simplest possible
example consider a wave function of two (spinless) fermions,
\begin{equation}
\Psi(\vec r_{1},\vec r_{2})=\frac{1}{\sqrt{2}}
\left[\phi(\vec r_{1})\chi(\vec r_{2})-\phi(\vec r_{2})\chi(\vec r_{1})
\right]
\label{fermwv}
\end{equation}
with two orthonormalized single-particle wave functions
$\phi(\vec r)$, $\chi(\vec r)$. Operator matrix elements between such
single Slater
determinants contain terms due to the antisymmetrization of coordinates
(``exchange contributions'' in the language of Hartree-Fock theory).
However, if the moduli of $\phi(\vec r)$, $\chi(\vec r)$ have only
vanishingly small overlap, these exchange correlations will also tend
to zero for any physically meaningful operator. This situation is generically
realized if the supports of the single-particle wave functions are essentially
centered around locations being sufficiently apart from each other, or the
particles are separated by a sufficiently large energy barrier.
In this case the
antisymmetrization present in Eq.~(\ref{fermwv}) has no physical effect.

Such observations clearly
justify the treatment of indistinguishable particles separated by macroscopic
distances as effectively distinguishable objects. So far, research in
Quantum Information Theory has concentrated on this case, where the exchange
statistics of particles forming quantum registers could be neglected, or was
not specified at all.

The situation is different if the particles constituting, say, qubits are
close together and possibly coupled in some computational process.
This the case for all proposals of quantum information
processing based on quantum dots technology
\cite{Loss98,Burkard99,Hu00,Schliemann01a,Friesen02,Vandersypen02}.
Here qubits are
realized by the spins of electrons living in a system of quantum dots.
The electrons have the possibility of tunneling eventually from one dot to the
other with a probability which can be modified by varying external parameters
such as gate voltages and magnetic field. In such a situation the fermionic
statistics of electrons is clearly essential.

Additional correlations in many-fermion-systems
arise if more than one Slater determinant  is involved, i.e. if
there is no single-particle basis such that a
given state of $N$ indistinguishable fermions can be
represented  as an elementary  Slater determinant (i.e. fully antisymmetric
combination of $N$ orthogonal single-particle states).
These correlations are  the analog of quantum entanglement in
separated systems and are essential for
quantum information processing in non-separated systems.

As an example consider a ``swap'' process, as discussed in the previous
subsection,exchanging the spin states of
electrons on coupled quantum dots by gating the tunneling amplitude between
them \cite{Burkard99,Schliemann01a}.
Before the gate is turned on, the two electrons in
the neighboring quantum dots are in a state represented by
a simple Slater determinant, and can be regarded as distinguishable since they
are separated by a large energy barrier. When the barrier is lowered,
more complex correlations between the electrons due to the dynamics arise.
Interestingly, as shown in Refs. \cite{Burkard99,Schliemann01a},
during such a process the system must necessarily enter a highly
correlated  state that cannot be represented by a single Slater determinant.
The final state of the gate operation, however,
is, similarly as the initial
one, essentially given by a single Slater determinant.
Moreover, by adjusting the gating time
appropriately one can also perform a ``square root of a swap'' which turns
a single  Slater determinant into a ``maximally'' correlated state
in much the same way \cite{Schliemann01a}. Illustrative details of
these processes will be given in section \ref{dynamics}.
In the end of such a process the electrons
can again be viewed as effectively distinguishable, but are in
a maximally entangled state in the usual sense of
distinguishable separated particles. In this sense the highly correlated
intermediate state can be viewed as a resource for the production of
entangled states.

In the following we give an elementary introduction to recent results
in the theory of quantum correlations in systems of indistinguishable
particles \cite{Schliemann01a,Schliemann01b,Eckert02}. This issue has
attracted recently considerable interest and intense discussions
\cite{Paskauskas01,Li01,Zanardi01,Zanardi02,Omar02,Gittings02,Ghirardi02,Zeng02,Wiseman02,Martin-Delgado02}.
The correlations to be discussed below are analogues of entanglement between
distinguishable parties.
However, to avoid confusion with the existing literature and in accordance
with Refs. \cite{Schliemann01b,Eckert02,Paskauskas01},
we shall reserve in the following the term ``entanglement'' for separated
systems and characterize the analogous quantum correlation phenomenon in
nonseparated systems in terms of the
Slater rank and the correlation measure
to be defined below. The results and concepts presented below
and in the references given are in various respect complementary to a
recent approach by Zanardi \cite{Zanardi01,Zanardi02}
stressing the dependence of
the notion of entanglement of the way a certain system is split into
subsystems.

For the purposes of this article we shall concentrate on elementary
results for the case of pure states of
two identical fermions. Results for mixed states and more than two fermions
can be found in \cite{Schliemann01a,Eckert02}. Results for the case
of identical bosons can be found in \cite{Li01,Paskauskas01,Eckert02}

We consider the case of two identical fermions sharing an
$n$-dimensional single-particle space ${\cal H}_{n}$ resulting in a
total Hilbert space ${\cal A}({\cal H}_{n}\otimes{\cal H}_{n})$ with
${\cal A}$ denoting the antisymmetrization operator. A general state vector
can be written as
\begin{equation}
|w\rangle=\sum_{a,b=1}^{n}w_{ab}f^{+}_{a}f^{+}_{b}|0\rangle
\label{defstate}
\end{equation}
with fermionic creation operators $f^{+}_{a}$ acting on the vacuum
$|0\rangle$. The antisymmetric coefficient
matrix $w_{ab}$ fulfills the normalization condition
\begin{equation}
{\rm tr}\left(\bar w w\right)=-\frac{1}{2}\,,
\end{equation}
where the bar stands for complex conjugation. Under a unitary transformation
of the single-particle space,
\begin{equation}
f^{+}_{a}\mapsto{\cal U}f^{+}_{a}{\cal U}^{+}=U_{ba}f^{+}_{b}\,,
\end{equation}
$w$ transforms as
\begin{equation}
w\mapsto UwU^{T}\,,
\end{equation}
where $U^{T}$ is the transpose (not the adjoint) of $U$. For any complex
antisymmetric matrix $n\times n$ matrix $w$ there is a unitary transformation
$U$ such that $w'=UwU^{T}$ has nonzero entries only in $2\times 2$
blocks along the diagonal \cite{Schliemann01b,Mehta77}. That is,
\begin{equation}
w'={\rm diag}\left[Z_{1},\dots,Z_{r},Z_{0}\right]
\quad{\rm with}\quad Z_i=\left[
\begin{array}{cc}
0 & z_i \\
-z_i & 0
\end{array}
\right]\,,
\label{z}
\end{equation}
$z_{i}\neq 0$ for $i\in\{1,\dots,r\}$, and $Z_{0}$ being the
$(n-2r)\times(n-2r)$ null matrix.
Each $2\times 2$ block $Z_{i}$ corresponds to an elementary Slater
determinant in
the state $|w'\rangle$. Such elementary Slater determinants are the analogues
of product states in systems consisting of distinguishable parties.
Thus, when expressed in such a basis, the state $|w\rangle$
is s sum of elementary Slater determinants
where each single-particle basis state enters not more
than one term. This property is analogous to the biorthogonality of the
Schmidt decomposition discussed above.
The matrix (\ref{z}) represents an expansion of $|w\rangle$ into a basis of
elementary Slater determinants with a minimum number $r$ of non-vanishing
terms. This number is analogous to the Schmidt rank for the distinguishable
case. Therefore we shall call it the {\em (fermionic) Slater rank} of
$|w\rangle$ \cite{Schliemann01b}, and an expansion of the above form a
{\em Slater decomposition} of $|w\rangle$.

We now turn to the case of two fermions in a four-dimensional single-particle
space. This case is realized in a system of two coupled quantum dots
hosting in total two electrons which are restricted to the lowest
orbital state on each dot. In such a system, a simple correlation
measure can be defined as follows \cite{Schliemann01a,Schliemann01b,Eckert02}:
For a given state (\ref{defstate}) with a coefficient matrix
$\omega_{ab}$ one defines a dual state $|\tilde\omega_{ab}\rangle$
characterized by the dual matrix
\begin{equation}
\tilde w_{ab}=\frac{1}{2}\sum_{c,d=1}^{4}\varepsilon^{abcd}\bar w_{cd}\,,
\label{defdual}
\end{equation}
with $\epsilon^{abcd}$ being the usual totally antisymmetric unit tensor.
Then the {\em correlation measure}
$\eta(|w\rangle)$ can be defined as
\footnote{The notation $\eta$ for the correlation measure was first
used in Refs.~\cite{Schliemann01a,Schliemann01b}. In Ref.~\cite{Eckert02}
the notation $\cal C$ was used in order to stress the analogy to the
``concurrence `` introduced by Wootters \cite{Wootters98}.}
\begin{equation}
\eta(|w\rangle)=\left|\langle\tilde w|w\rangle\right|
=\left|\sum_{a,b,c,d=1}^{4}
\varepsilon^{abcd}w_{ab}w_{cd}\right|
=\left|8\left(w_{12}w_{34}+w_{13}w_{42}+w_{14}w_{23}\right)\right|\,.
\end{equation}
Obviously, $\eta(|w\rangle)$ ranges from zero to one. Importantly it vanishes
if and only if the state $|w\rangle$ has the fermionic Slater rank one, i.e.
$\eta(|w\rangle)$
is an elementary Slater determinant. This statement was proved first in
Ref.~\cite{Schliemann01a}; an alternative proof can be given using the
Slater decomposition of $|w\rangle$ and observing that
\begin{equation}
\det w=\left(\frac{1}{8}\langle\tilde w|w\rangle\right)^{2}\,.
\label{det}
\end{equation}
This is just a special case of a general relation expressing the
determinant of an antisymmetric $(2K)\times(2K)$ matrix $w$ by
the square of its Pfaffian,
\begin{equation}\label{eqn:asdet}
\det w=\Big(\frac{1}{2^{K}K!}\sum_{i_{1},\dots,i_{2K}=1}^{2K}
\varepsilon^{i_{1}\dots i_{2K}}w_{i_{1}i_{2}}\dots w_{i_{2K-1}i_{2K}}
\Big)^{2}\,.
\end{equation}

The quantity $\eta(|w\rangle)$ measures quantum correlation contained
in the two-fermion state $|w\rangle$ which are beyond simple
antisymmetrization effects. This correlation measure in under many aspects
analogous to the entanglement measure ``concurrence'' used in systems
of two distinguishable qubits \cite{Wootters98}. These analogies are
discussed in detail in \cite{Eckert02} including also the case of
indistinguishable bosons. An important difference between just
two qubits, i.e. two distinguishable two-level systems,
and the present case of two electrons in a two-dot system is
that in latter system both electrons can possibly occupy the same dot
while the other is empty. Therefore the total Hilbert space is larger
than in the two-qubit system, and a generalized correlation measure
becomes necessary. Furthermore, similar as in the two-qubit case, the
correlation measure $\eta$ defined here for pure states of two fermions
has a natural extension to mixed fermionic and bosonic states
\cite{Schliemann01b,Eckert02}.

A convenient choice to make contact between the general state labels
$a,b,\dots\in\{1,2,3,4\}$ used here and the basis states of the
preceding section is given by $(1,2,3,4)=
(A\uparrow,A\downarrow,B\uparrow,B\downarrow)$.
With this convention, a state vector spanned by $|S_{2}\rangle$ and
$|S_{3}\rangle$ only has $w_{12}$ and $w_{34}$ as its only independent
non-zero coefficients in $w$. Such a state lies fully in the subspace
of double occupancies, and its entanglement is purely due to the
orbital degrees of freedom:
\begin{equation}
\eta_{orb}=8|w_{12}w_{34}|\,.
\label{orbit}
\end{equation}
On the other hand, a state spanned by $|S_{1}\rangle$ and $|T^{0}\rangle$
has no double occupancies and
is entangled purely with respect to the spin degrees of freedom,
\begin{equation}
\eta_{spin}=8|w_{14}w_{23}|\,.
\label{spin}
\end{equation}
For a general state vector, both kinds of correlations (orbital and
spin) contribute to
$\eta(w)$.

A direct connection of the degree of quantum correlation $\eta$
in the singlet ground state of the double dot system and its
tunneling amplitude was found recently in \cite{Golovach02}.
The ground state reads
\begin{equation}
|w_{0}\rangle=\frac{1}{\sqrt{2(1+\phi^{2})}}
\left((1+\phi)|S_{1}\rangle+(1-\phi)|S_{2}\rangle\right)
\end{equation}
with an energy given in (\ref{stsplitting}). Here the
``interaction parameter''
\begin{equation}
\phi=\sqrt{1+\left(\frac{4 t_{H}}{U_{H}}\right)^{2}}-\frac{4 t_{H}}{U_{H}}\,.
\end{equation}
has been introduced, and the
correlation measure of the ground state is given by
\begin{equation}
\eta=\frac{2\phi}{1+\phi^{2}}
\end{equation}
At zero tunneling $t_{H}=0$ the ground state singlet does not contain any
double occupancies and is degenerate with the triplet. At large tunneling
$t_{H}\gg U_{H}$ both electrons are in symmetric orbital states differing
in spin. This state is a single Slater determinant, and the correlation
measure is zero.


\section{Dynamics of entanglement in quantum gate operations}
\label{dynamics}

We now continue with our investigation of the dynamics of the double
quantum
dot qubit swapping process generated by a time-dependent tunneling
amplitude.

Let us first consider the case of inversion symmetry
along the axis connecting the dots.
As explained in Sec. \ref{model} this problem
can be reduced essentially to the time evolution of a
pseudospin-$\frac{1}{2}$-object in a magnetic field having a time-dependent
component in the $x$-direction of the pseudospin space. In the course of swapping,
the triplet contribution to the incoming state (\ref{instate}) will
just pick up a phase factor according to its constant eigenvalue, while the
singlet contribution will mix with the other singlet $|S_{2}\rangle$.
Therefore, a finite probability for double occupancies will necessarily
arise {\em during} the swapping process. However, if these amplitudes can be
suppressed sufficiently when the swapping is complete (as in
the adiabatic limit), errors in the quantum computation
can be avoided. Thus we are left with the question of how close the dynamics
of our formal spin-$\frac{1}{2}$-object is to its adiabatic limit.
We note that, in the adiabatic limit, no Berry phase
occurs in the time evolution of the singlet states,
since the motion of the formal spin is restricted to a plane. Hence
the solid angle encircled in a round trip is strictly zero.

The integration of the Schr\"odinger equation for our time-dependent two-level
problem is in general non-elementary. However, there is a considerable body of
literature, starting with early work by Landau \cite{Landau32}, Zener
\cite{Zener32},
and Rosen and Zener \cite{Rosen32}, where particular cases of this
problem were reduced to
well-known differential equation of mathematical physics such as the
hypergeometric equation.
This work was reviewed and generalized very recently in
\cite{Ishkhanyan00}. However, such an approach still works only for special
time-dependent Hamiltonians, i.e., in the present context, only for special
shapes of
the tunneling pulse $t_{H}(t)$, and many quantities of
interest are given by complicated non-elementary
expressions which require numerical evaluation. For this reasons, and for the
sake of brevity of our presentation, we shall resort to numerical integrations
of the Schr\"odinger equation. From such studies we will see that the
range of adiabaticity is remarkably large.
Our numerical findings will be corroborated and made physically plausible
by well-known analytical results for Landau-Zener-type transitions in
simplified cases.

To be specific, we consider a time-dependent tunneling of the form
\begin{equation}
t_{H}(t)=\frac{\Delta}
{1+\frac{\cosh\left(t/\tau\right)}{\cosh\left(T/(2\tau)\right)}}\,.
\label{tunnel}
\end{equation}
This is a tunneling pulse which is switched on and off exponentially with
a characteristic time $\tau$. It has a duration of $T$ and an amplitude
given by $\Delta$ (for $T\gg\tau$). Therefore this form is flexible enough
to describe the essential features of a pulse.
The exponential switching is motivated
by the exponential-like dependence of the tunneling matrix element on external
parameters \cite{Burkard99}.

A typical situation is shown in figure \ref{fig1} for a switching time
of $\tau=4\hbar/U_{H}$, an amplitude of $\Delta=U_{H}/8$ and the duration
$T$ adjusted to enable  single swap operation. The figure shows the results
of a numerical integration of the time-dependent Schr\"odinger equation
using the fourth order Runge-Kutta scheme. The time-dependent
tunneling amplitude $t_{H}(t)$ is plotted (in units $U_{H}$) as a dotted line.
The square amplitude of
the incoming state (\ref{instate}) and the outgoing state (\ref{outstate}) are
shown as thick lines. The square amplitudes of the singlets
$|S_{1}\rangle$ and $|S_{2}\rangle$ are denoted by $|\varphi_{1}|^{2}$ and
$|\varphi_{2}|^{2}$, respectively, and plotted as long-dashed lines.
The probability of double occupancies is given by $|\varphi_{2}|^{2}$.
As one can see from the figure, this quanity is finite during the
swapping process but strongly suppressed afterwards. The measure of
entanglement $\eta(t)$ is also shown in the figure. It is zero for the
non-entangled incoming and outgoing state, and achieves its
maximum value of almost unity in the middle of the process. This quantifies and
shows explicitly the entanglement of the quantum state during the
swapping process.

The probability $|\varphi_{2}|^{2}$ for double occupancy after switching off
the tunneling depends on the switching time $\tau$, the amplitude $\Delta$ and
also on the duration $T$ of the tunneling pulse, i.e. on the exact time when
the switching off sets in. However, our numerics suggest that there is
an upper bound for $|\varphi_{2}|^{2}$ at given $\tau$ and $\Delta$.
In the above example the double occupancy probability after the swapping
process is smaller than $10^{-10}$, which is a very tiny value.
A typical order of
magnitude for the double occupancy probability is $10^{-6}$ for amplitudes
$\Delta<U_{H}$ and switching times $\tau>4\hbar/U_{H}$. In fact, also
larger
values of $\Delta$ (being still comparable with $U_{H}$) can be possible,
leading to double occupancy probabilities of the same order,
while this probability significantly increases if $\tau$
becomes smaller than $4\hbar/U_{H}$. Thus, this value characterizes the
region where the motion of the system is close to its adiabatic limit and
is remarkably small on the natural time scale of the system given by
$\hbar/U_{H}$, while adiabatic behavior is in general expected for a
particularly slow time evolution.

This large range of quasi-adiabatic behavior can be understood qualitatively
by considering a simplified situation where the tunneling is switched on and
off linearly in time and is constant otherwise. Then, non-adiabatic effects
can occur only during the sharply defined switching processes. For simplicity,
we consider the first switching process only where the tunneling has
the time dependence $t_{H}=(\Delta/\tau)t$, $t\in[0,\tau]$. To enable
analytical progress let us further assume $t\in[-\infty,\infty]$, which should
lead to an upper bound for the probability of non-adiabatic transitions
due to the switching. This problem was considered a long time ago
by Landau \cite{Landau32} and by Zener \cite{Zener32}.
The result of reference \cite{Zener32} for the probability of non-adiabatic
transitions reads
\begin{equation}
P_{{\rm nad}}=e^{-\alpha}
\label{traprob}
\end{equation}
with an adiabaticity parameter
\begin{equation}
\alpha=\frac{\pi}{8}\frac{U_{H}^{2}}{\hbar(\Delta/\tau)}\,.
\label{adpar}
\end{equation}
We see
that the probability for non-adiabatic transitions is exponentially suppressed
with increasing switching time $\tau$. This exponential dependence explains
qualitatively the above observation of a large range of quasi-adiabatic
behavior. To obtain an estimate for a nonlinear switching one may replace
the ratio $(\Delta/\tau)$ in the denominator of (\ref{adpar}) by the maximum
time derivative of the tunneling $t_{H}(t)$ (giving
$\alpha=\pi U_{H}^{2}/3\hbar(\Delta/\tau)$ for the pulse (\ref{tunnel})).

A similar exponential dependence of the  probability for
non-adiabatic transitions on the switching time $\tau$ was also found
analytically by Rosen and Zener \cite{Rosen32} for a particular
two-parametric pulse of the form
\begin{equation}
t_{H}(t)=\Delta/\cosh(t/\tau)\,.
\end{equation}
In this case non-adiabatic transitions occur with a probability
\begin{equation}
P_{{\rm nad}}=\sin^{2}\left(\Delta\tau/(2\hbar)\right)/
\cosh^{2}\left(U_{H}\tau/(2\hbar)\right)\,.
\label{traprob2}
\end{equation}
To illustrate the behavior in the strongly non-adiabatic case we have plotted
in figure \ref{fig2} $|\varphi_{1}|^{2}$ and
$|\varphi_{2}|^{2}$ for the same situation as in figure \ref{fig1} but with a
four times smaller
ramp time of only $\tau=\hbar/U_{H}$. In this case small oscillations occur in
the time-evolution of these two quantities during the tunneling pulse,
which can be understood in terms
of the eigenspectrum at a given tunneling $t_{H}=\Delta$. Additionally, a
sizable double occupancy probability of about $0.005$ remains after the pulse,
as shown in the inset.

Figure \ref{fig3} shows a square root of a swap, which is obtained from the
situation of figure \ref{fig1} by halfing the duration $T$ of the tunneling
pulse. The resulting state is a fully entangled complex
linear combination of the
states $|S_{1}\rangle$ and $|T^{0}\rangle$, or, equivalently, of the
incoming state (\ref{instate}) and the outgoing state (\ref{outstate}) of the
full swap. Again, the weight of the
doubly occupied state $|S_{2}\rangle$ is strongly suppressed after
the tunneling pulse. As a consequence, Eq. (\ref{orbit}) implies that
$\eta_{orb}=0$ after completion of switching, while
$\eta=\eta_{spin}=8|w_{14}w_{23}|=1$. This shows that the entanglement
of the two electrons is entirely in the spin (and not in the orbital)
degrees of freedom after switching.

Let us finally consider swapping processes when the inversion symmetry along
the axis connecting the dots is broken. Such processes are governed by the
Hamiltonian (\ref{3ham}) in the presence of a finite matrix element $F$.
Our numerical results are in this case qualitatively the same as before
with the admissible switching times $\tau$ slightly growing with increasing
$F$. In figure \ref{fig4} we illustrate our findings for a comparatively large
off-diagonal element $F=0.4U_{H}$. The additional Coulomb matrix element is
$X=0.2U_{H}$, and the parameters of the tunneling pulse are
$\tau=8\hbar/U_{H}$ and $\Delta=U_{H}/8$ with a duration $T$ appropriate
for a single swapping. As a result, a clean swapping operation can be
performed also in the absence of inversion symmetry.

We note that the Hund-Mulliken scheme used here is restricted to the
low-energy  sector where only the lowest single-particle energy
levels (with typical spacings $\delta\epsilon$) are kept.
For this scheme to be valid also in a switching process, we need to require
that time-dependent changes must  be performed adiabatically also with respect to
the time scale set by
$\hbar/\delta\epsilon$, i.e. we need $\tau >
\hbar/\delta\epsilon$\cite{Burkard99}. On the other hand,
to suppress double occupancy errors we have seen that
the adiabaticity parameter $\alpha$ of Eq.
(\ref{adpar}) must exceed unity, implying that
$\tau > 8\hbar\Delta/ (\pi U_H^2)$.
Thus, the adiabaticity condition for switching becomes more generally,
\begin{equation}
\tau> \tau_{min}:= \max
\left\{ {\hbar\over \delta\epsilon}, {8\hbar\over \pi}
{\Delta
\over U_H^2}\right\}\, .
\end{equation}
There are now two particular cases we can distinguish. First, if the
effective Coulomb charging energy exceeds the level spacing,
i.e. $U_H>\delta\epsilon$,
we obtain $\tau_{min}=\hbar/\delta\epsilon$,
since for consistency we have $\Delta <\delta\epsilon$. Thus, when the
switching is adiabatic with respect to the scale set by $\delta\epsilon$,
errors due to double occupancy are automatically excluded.
In the second case with $U_H < \sqrt{\Delta \delta\epsilon}<\delta\epsilon$
(``ultrasmall quantum dots"),
we obtain $\tau_{min}=8\hbar\Delta/(\pi U_H^2)$, which means that the
overall  condition for adiabaticity is determined by the no-double occupancy
criterion.  Using typical material parameters for GaAs quantum
dots\cite{Kouwenhoven97}, we can estimate\cite{Burkard99} that $\tau_{min}$ is of
the order of 50 ps.


\section{Summary}
\label{summary}

We have given an overlook on the physics of gate operations between
quantum dot spin-qubits emphasizing the dynamics of quantum entanglement
between such objects. For these purposes, the low-energy physics of a
lateral pair of identical quantum dots can be effectively reduced to a
(time-dependent) two-state problem,
In particular we have analyzed the swap and
``square root of a swap'' operations by means of numerical simulations
of the time-dependent Schr\"odinger equation. One of the main conclusions
is that double occupancy error will not be a fatal problem for the
operation of such quantum gates. This is due to a surprisingly
large range of the quasiadiabatic regime in the quantum dynamics of the
aforementioned two-state system
Our numerical results are
corroborated and physically made plausible by analytical results
on related Landau-Zener-type transitions.
We have also reported on recent results on electron spin decoherence
in semiconductor nanostructures due to hyperfine interaction with surrounding
nuclear spins.

Finally, the indistinguishable character of the electrons involved in the
realization of quantum dot spin-qubits leads to the interesting question
of entanglement-analogous quantum correlations between
{\em indistinguishable} particles. Here we have provided an introduction
to this issue, which has been addressed in the
literature since only very recently and constitutes a new direction of
work in quantum information theory.


%
\begin{figure}
\centerline{\includegraphics[width=12cm]{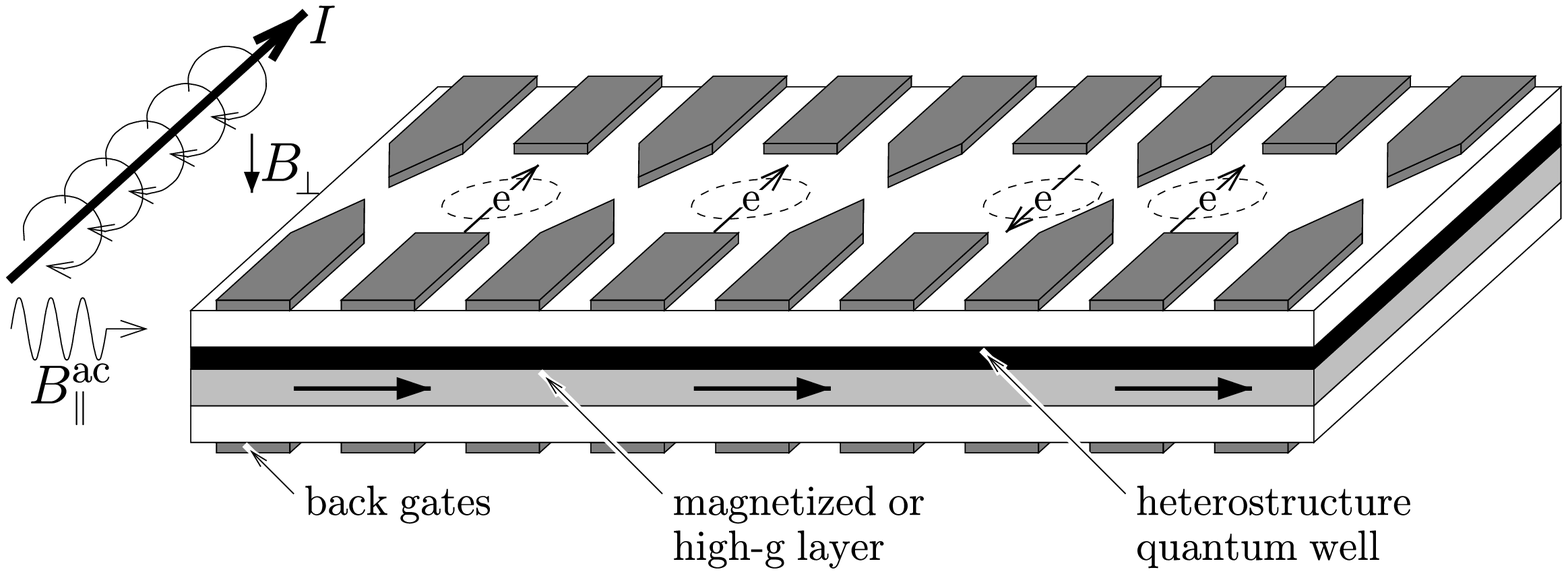}}
\caption{\label{figArray}
Quantum dot array, controlled by electrical gating.
The electrodes (dark gray) define quantum dots (circles) by confining
electrons. Additional gates and (dc and ac) external magnetic fields
enable single-qubit rottaions. Two-qubit operations are achived by
gating the effective tunneling amplitude between neighboring quantum dots.
This introduces an effective time-dependent exchange coupling between
the electron spins.}
\end{figure}

\begin{figure}
\centerline{\includegraphics[width=8cm]{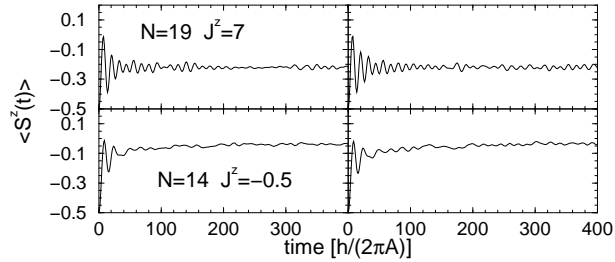}}
\caption{Upper
panels: $\langle S^{z}(t)\rangle$ for a system of size $N=19$
being initially in a  randomly correlated nuclear spin state in
the subspace with $J^{z}=7$ where $\vec J=\vec S+\sum_{i}\vec I_{i}$
denotes the total spin of the system.
The two panels represent two
different randomly chosen initial conditions. Lower panels:
Analogous data for $N=14$ and a completely unpolarized nuclear
spin system ($J^{z}=-1/2$). In both cases the simulation data does
practically not depend on the initial condition. \label{nuclearfig1}}
\end{figure}
\begin{figure}
\centerline{\includegraphics[width=8cm]{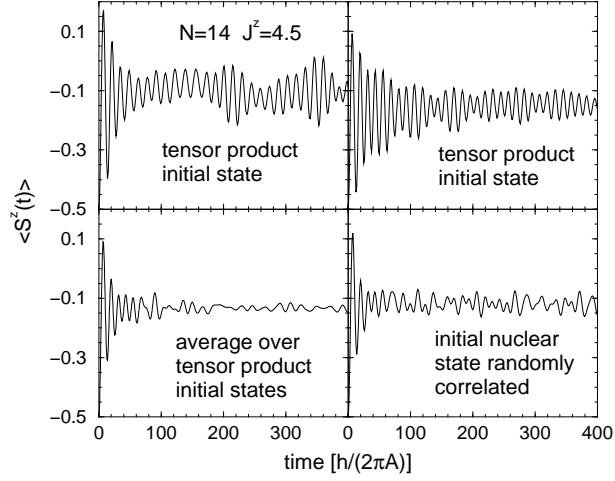}}
\caption{Upper
panels: time evolution of the electron spin $\langle
S^{z}(t)\rangle$ for a system with 14 nuclear spins being
initially in an uncorrelated tensor product state in the subspace
 $J^{z}=9/2$. The time evolution of the electron spin depends strongly
on the initial state of the nuclear spins.
Lower left panel: data
of the same type as above but averaged over all possible
uncorrelated initial states with $J^{z}=9/2$. Lower right panel:
$\langle S^{z}(t)\rangle$ for the same system being initially in a
randomly chosen correlated state. This time evolution closely mimics
the avergaged data of the lower left panel.
\label{nuclearfig2}}
\end{figure}
\begin{figure}
\centerline{\includegraphics[width=8cm]{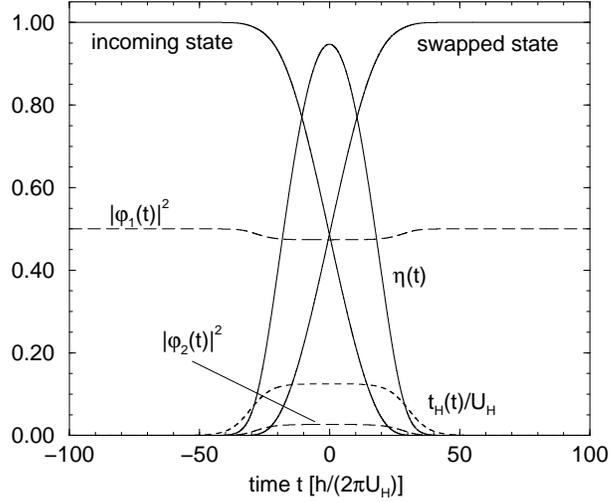}}
\caption{A swap process as a function of time. The
tunneling amplitude $t_{H}(t)$ is plotted (in units of $U_{H}$) as a dotted
line.
The square amplitude of the incoming state (\protect{\ref{instate}})
and the outgoing state (\protect{\ref{outstate}}) are
shown as thick lines. The square amplitudes of the singlets
$|S_{1}\rangle$ and $|S_{2}\rangle$ are denoted by $|\varphi_{1}|^{2}$ and
$|\varphi_{2}|^{2}$, respectively, and plotted as long-dashed lines.
The measure of entanglement $\eta(t)$ is also shown.
\label{fig1}}
\end{figure}
\begin{figure}
\centerline{\includegraphics[width=8cm]{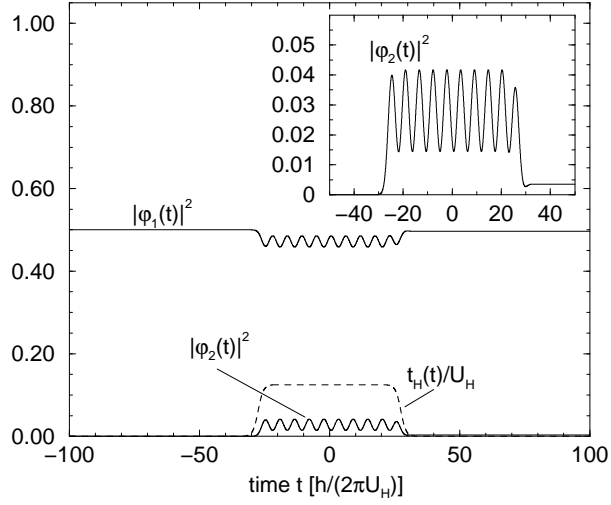}}
\caption{The square amplitudes of the singlet states $|S_{1}\rangle$ and
$|S_{2}\rangle$ for the same situation as in figure \protect{\ref{fig1}}, but
with a four times smaller ramp time of only $\tau=\hbar/U_{H}$.
The inset shows
$|\varphi_{2}(t)|^{2}$ on a magnified scale. The dynamics of the system is
clearly in the non-adiabatic regime.
\label{fig2}}
\end{figure}
\begin{figure}
\centerline{\includegraphics[width=8cm]{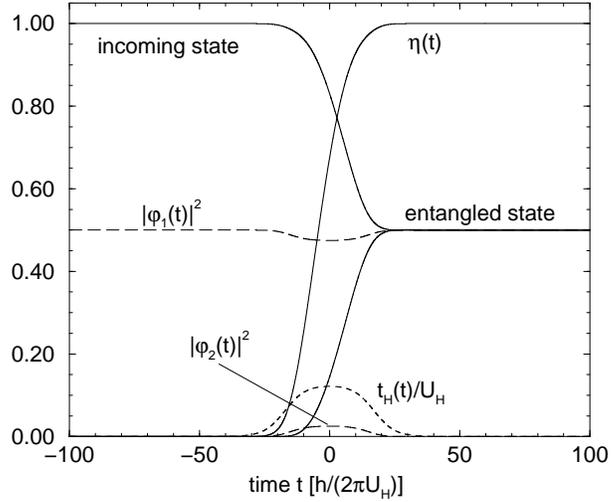}}
\caption{A square root of a swap, which is obtained from the situation of
figure \protect{\ref{fig1}} by halfing the pulse duration $T$. The probability
of double occupancies is again strongly suppressed after the tunneling pulse.
The resulting state is a fully entangled complex linear combination of
$|S_{1}\rangle$ and $|T^{0}\rangle$, or, equivalently, of the
incoming state (\protect{\ref{instate}}) and the outgoing state
(\protect{\ref{outstate}}) of the full swap. The quantum mechanical weigths
of the latter states are plotted as thick solid lines
\label{fig3}}
\end{figure}
\begin{figure}
\centerline{\includegraphics[width=8cm]{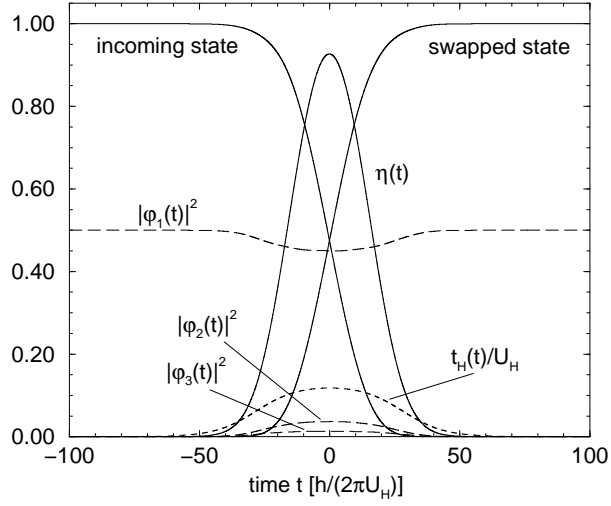}}
\caption{A swapping processes in the absence of inversion symmetry along
the axis connecting the dots. The square amplitudes of the singlet states
$|S_{i}\rangle$, $i\in\{1,2,3\}$, are denoted by $|\varphi_{i}|^{2}$.
The additional matrix elements entering the
Hamiltonian (\ref{3ham}) are $X=0.2U_{H}$ and $F=0.4U_{H}$.
The parameters of the tunneling pulse are
$\tau=8\hbar/U_{H}$ and $\Delta=U_{H}/8$ with a duration $T$ appropriate
for a single swapping. As a result, a clean swapping operation can be
performed also in the absence of inversion symmetry.
\label{fig4}}
\end{figure}
\end{document}